\documentclass[12pt]{iopart}
\usepackage{graphicx}
\usepackage{bm}

\begin{document}

\title[Gravastars supported by nonlinear electrodynamics]
{Gravastars supported by nonlinear electrodynamics}

\author{Francisco S. N. Lobo\footnote[2]{flobo@cosmo.fis.fc.ul.pt}}
\address{Centro de Astronomia
e Astrof\'{\i}sica da Universidade de Lisboa,\\
Campo Grande, Ed. C8 1749-016 Lisboa, Portugal, and}
\address{Institute of Cosmology \& Gravitation,
University of Portsmouth,\\
Portsmouth, PO1 2EG, United Kingdom}

\author{Aar\'{o}n V. B. Arellano\footnote[1]{vynzds@yahoo.com.mx}}
\address{Facultad de Ciencias,
Universidad Aut\'{o}noma del Estado de M\'{e}xico, \\
El Cerrillo, Piedras Blancas, C.P. 50200, Toluca, M\'{e}xico}
%


\begin{abstract}

Gravastar models have recently been proposed as an alternative to
black holes, mainly to avoid the problematic issues associated
with event horizons and singularities. In this work, a wide
variety of gravastar models within the context of nonlinear
electrodynamics are constructed. Using the $F$ representation,
specific forms of Lagrangians are considered describing magnetic
gravastars, which may be interpreted as self-gravitating magnetic
monopoles with charge $g$. Using the dual $P$ formulation of
nonlinear electrodynamics, electric gravastar models are
constructed by considering specific structural functions, and the
characteristics and physical properties of the solutions are
further explored. These interior nonlinear electrodynamic
geometries are matched to an exterior Schwarzschild spacetime at a
junction interface.

\end{abstract}


\maketitle

\section{Introduction}

The generally well-accepted notion of a black hole is deeply
rooted in the general relativist community, although a
considerable number of particle and condensed matter physicists
view the respective notion of event horizons with some suspicion
and alarm \cite{Laughlin,Mazur,Chapline}. In addition to this, one
still encounters the existence of misconceptions and a certain
ambiguity inherent to the Schwarzschild solution in the literature
(see Ref. \cite{Doran} for a review). Despite the fact that the
evidence of the existence of black holes in the Universe is very
convincing, a certain amount of scepticism regarding the
observational data is still encountered \cite{AKL}. This
scepticism has inspired new and fascinating ideas. In particular,
models replacing the interior Schwarzschild solution with compact
objects, and thus, doing away with the problems of the singularity
at the origin and the event horizon, have been proposed to some
extent in the literature. However, we do emphasize that the
interior structure of realistic black holes has not been
satisfactorily determined, and is still open to considerable
debate.
In this line of thought, an interesting alternative to black holes
has recently been proposed, namely, the ``gravastar'' ({\it
grav}itational {\it va}cuum {\it star}) picture developed by Mazur
and Mottola \cite{Mazur}. In this model, and in the related
picture developed by Laughlin {\it et al} \cite{Laughlin}, the
quantum vacuum undergoes a phase transition at or near the
location where the event horizon is expected to form. The
Mazur-Mottola model is constituted by an onion-like structure with
five layers, including two thin-shells, with surface stresses
$\sigma_\pm$ and ${\cal P}_\pm$, where $\sigma$ is the surface
energy density and ${\cal P}$ is the surface tangential pressure.
The interior of the solution is replaced by a segment of de Sitter
space, which is then matched to a finite thickness shell of stiff
matter with the equation of state $p=\rho$. The latter is then
matched to an external Schwarzschild vacuum with $p=\rho=0$.
Related models, analyzed in a different context have also been
considered by Dymnikova \cite{Dymnikova}.

Although Mazur and Mottola argued for the thermodynamic stability
of their configuration, the full dynamic stability against
spherically symmetric perturbations was carried out in Ref.
\cite{VW}, through a simplified construction of the Mazur-Mottola
model. The configuration was simplified to a three-layer solution,
i.e., a de Sitter interior solution was matched to a Schwarzschild
exterior solution at a junction surface, comprising of a thin
shell with surface stresses $\sigma$ and ${\cal P}$. It was found
that many equations of state exist that imply the dynamical
stability of the gravastar configurations. The latter dynamical
stability was generalized to an anti-de Sitter or de Sitter
interior and a Schwarzschild-(anti)-de Sitter or
Reissner--Nordstr\"{o}m exterior \cite{Carter}. In Ref.
\cite{CFV}, by disregarding the presence of thin shells, a
generalized class of gravastar models was analyzed. It was found,
using these models, that gravastars cannot be perfect fluid
spheres, and that they necessarily exhibit anisotropic pressures.
Specific stable solutions, with respect to axial perturbations,
possessing continuous pressures were also further explored
\cite{DeBen}. In this context, one may also replace the interior
de Sitter spacetime with a solution governed by the dark energy
equation of state, $\omega=p/\rho$, where $\omega <-1/3$
\cite{darkstar}. Note that the particular case of $\omega=-1$
reduces to the de Sitter solution, and for $\omega<-1$, one ends
up with phantom energy, an exotic cosmic fluid which violates the
null energy condition. Now, the notion of dark energy is that of a
spatially homogeneous cosmic fluid, however, inhomogeneities may
arise through gravitational instabilities. Thus, these
inhomogeneous solutions, denoted as {\it dark energy stars}
\cite{Chapline,darkstar}, may possibly originate from density
fluctuations in the cosmological background. Note that now the
pressure in the equation of state $p=\omega \rho$ is a radial
pressure, and the transverse pressure may be determined from the
field equations, as emphasized in Refs. \cite{phantom}, in a
rather different context.

In Ref. \cite{Bilic}, motivated by low energy string theory, an
alternative model was constructed by replacing the de Sitter
regime with an interior solution governed by a Chaplygin gas
equation of state, interpreted as a Born-Infeld phantom gravastar.
In this context, the extension of the latter Born-Infeld gravastar
variation to rather more general nonlinear electrodynamic models
shall prove extremely interesting, and this is, in fact the
analysis of the present paper. It is interesting to note that
nonlinear electrodynamics has recently been revived, mainly due to
the fact that these models appear as effective theories at
different levels of string/M-theory, in particular, in D$p-$branes
and supersymmetric extensions, and non-Abelian generalizations (we
refer the reader to Ref. \cite{Witten} for a review), and have
enjoyed a wide variety of applicability, namely, in cosmology
\cite{NLED-cosmology}.
Traversable wormholes have also been explored in nonlinear
electrodynamics \cite{Arellano1,Arellano2}. It was found that
regular pure magnetic evolving wormholes exist \cite{Arellano1}.
However, static spherically symmetric and stationary axisymmetric
traversable wormholes have been ruled out, mainly due to the
presence of event horizons, the non-violation of the null energy
condition at the throat, and due to the imposition of the
principle of finiteness, which states that a satisfactory theory
should avoid physical quantities becoming infinite
\cite{Arellano2}. In fact, the Born-Infeld model was inspired
mainly on the principle of finiteness, in order to remedy the fact
that the standard picture of a point particle possesses an
infinite self-energy \cite{BI}. Later, Pleba\'{n}ski presented
other examples of nonlinear electrodynamic Lagrangians
\cite{Pleb}, and showed that the Born-Infeld theory satisfies
physically acceptable requirements.

It was in nonlinear electrodynamics that the first exact regular
black hole solutions to the Einstein field equation were found
\cite{Garcia,Garcia2,Garcia3}. In this context, magnetic black
holes and monopoles \cite{Bronnikov1}, regular electrically
charged black holes of a hybrid type containing a magnetically
charged core \cite{Burinskii}, and electrically charged structures
with a regular de Sitter center \cite{Dymnikova2}, have been
found. In this work, we are interested in the construction of a
wide variety of gravastar models coupled to nonlinear
electrodynamics. Using the $F$ representation, we consider
specific forms of Lagrangians describing magnetic gravastars,
which may be interpreted as self-gravitating magnetic monopoles
with a charge $g$. Using the $P$ formulation of nonlinear
electrodynamics, it is easier to find electric solutions, and
considering specific structural functions we further explore the
characteristics and physical properties of the solutions. We match
these interior nonlinear electrodynamic geometries to an exterior
Schwarzschild spacetime, thus avoiding the problematic issues
related to the singularities and event horizons.

This paper is organized in the following manner: In section
\ref{sec1}, we outline the field equations, and analyze some
characteristics of the solutions. In section \ref{sec2}, we
consider the $F$ representation of nonlinear electrodynamics, and
explore specific magnetic solutions in some detail. In section
\ref{sec3}, we outline the dual $P$ formalism and extensively
analyze electric gravastar solutions. In section
\ref{sec:conclusion}, we conclude.

\section{Field equations}\label{sec1}

We consider the action of $(3+1)-$dimensional general relativity
coupled to nonlinear electrodynamics given in the following form
\begin{equation}
S=\int \sqrt{-g}\left[\frac{R}{16\pi}+L(F)\right]\,d^4x  \,,
\end{equation}
where $R$ is the Ricci scalar, and $L(F)$ is a gauge-invariant
electromagnetic Lagrangian. The latter depends on a single
invariant $F$ given by $F=F^{\mu\nu}F_{\mu\nu}/4$ \cite{Pleb},
where the antisymmetric tensor
$F_{\mu\nu}=A_{\nu,\mu}-A_{\mu,\nu}$ is the electromagnetic field.
In Einstein-Maxwell theory, the Lagrangian takes the form $L(F)=
-F/4\pi$. However, we consider more general choices for the
electromagnetic Lagrangians. Note that the Lagrangian may also be
constructed using a second invariant $G \sim
F_{\mu\nu}{}^*F^{\mu\nu}$, where the asterisk $^*$ denotes the
Hodge dual with respect to $g_{\mu\nu}$. However, we only consider
$F$, as this provides solutions that are interesting enough.

Varying the action with respect to the gravitational field
provides the Einstein field equations $G_{\mu\nu}=8\pi
T_{\mu\nu}$, where $G_{\mu\nu}$ is the Einstein tensor, and the
stress-energy tensor, $T_{\mu\nu}$, is given by
\begin{equation}
T_{\mu\nu}=g_{\mu\nu}\,L(F)-F_{\mu\alpha}F_{\nu}{}^{\alpha}\,L_{F}\,.
    \label{4dim-stress-energy}
\end{equation}
with $L_F=dL/dF$.

The electromagnetic field equations are the following
\begin{equation}
\left(F^{\mu\nu}\,L_{F}\right)_{;\mu}=0 \;,  \qquad
\left(^*F^{\mu\nu}\right)_{;\mu}=0 \,.
     \label{em-field}
\end{equation}
The first equation may be obtained by varying the action with
respect to the electromagnetic potential $A_\mu$. The second
relationship, in turn, may be obtained from the Bianchi
identities.

Consider a static and spherically symmetric spacetime, in
curvature coordinates, given by the following line element
\begin{eqnarray}
ds^2&=&-e^{2\alpha(r)}\,dt^2 +e^{2\beta(r)}\,dr^2+r^2(d\theta
^2+\sin ^2{\theta}\, d\phi ^2)  \,, \label{generalmetric}
\end{eqnarray}
where $\alpha$ and $\beta$ are functions of $r$.

Taking into account the symmetries of the geometry, the non-zero
compatible terms for the electromagnetic tensor are
\begin{equation}
F_{\mu\nu}=2E(x^\alpha)\,\delta^{[t}_\mu
\,\delta^{r]}_\nu+2B(x^\alpha)\,\delta^{[\theta}_\mu
\,\delta^{\phi]}_\nu \label{em-tensor}\,,
\end{equation}
so that the only non-zero terms are $F_{tr}=E(x^\mu)$ and
$F_{\theta\phi}=B(x^\mu)$.

The Einstein tensor components, in an orthonormal reference frame,
(with $c=G=1$) are given by
\begin{eqnarray}
G_{\hat{t}\hat{t}}&=&\;\frac{e^{-2\beta}}{r^2}
\,\left(2\beta'r+e^{2\beta}-1 \right)  \label{rho}\,,\\
G_{\hat{r}\hat{r}}&=&\frac{e^{-2\beta}}{r^2}
\,\left(2\alpha'r-e^{2\beta}+1 \right) \label{pr}\,,\\
G_{\hat{\phi}\hat{\phi}}&=&G_{\hat{\theta}\hat{\theta}}=
\frac{e^{-2\beta}}{r}\,\left[-\beta'+\alpha'+r\alpha''
+r(\alpha')^2-r\alpha' \beta'\right] \label{pt}\,,
\end{eqnarray}
where the prime denotes a derivative with respect to the radial
coordinate, $r$. Now, it is a simple matter, using equation
(\ref{4dim-stress-energy}) in an orthonormal reference frame, to
verify that $T_{\hat{t}\hat{t}}=-T_{\hat{r}\hat{r}}$. Thus, using
the Einstein field equation, and taking into account equations
(\ref{rho}) and (\ref{pr}), we have $\alpha'+\beta'=0$, which
provides the solution $\alpha=-\beta+C$. The constant $C$ may be
absorbed by defining a new time coordinate, so that without a loss
of generality, we may consider $C=0$.

For notational and computational convenience, we consider the
metric fields in the following form
\begin{equation}
e^{2\alpha(r)}=e^{-2\beta(r)}=\left[1-\frac{2m(r)}{r}\right] \,,
\end{equation}
so that the Einstein tensor components reduce to
\begin{eqnarray}
G_{\hat{t}\hat{t}}=-G_{\hat{r}\hat{r}}=\frac{2m'}{r^2}\,, \qquad
G_{\hat{\phi}\hat{\phi}}=G_{\hat{\theta}\hat{\theta}}=
-\frac{m''}{r}\,.
\end{eqnarray}
The function $m(r)$ may be considered the quasi-local mass, and is
denoted as the mass function. The stress-energy tensor, in the
orthonormal reference frame, assumes the diagonal form
$T_{\hat{\mu}\hat{\nu}}={\rm diag}(\rho,p_r,p_t,p_t)$, in which
$\rho(r)$ is the energy density, $p_r(r)$ is the radial pressure,
and $p_t(r)$ is the lateral pressure measured in the orthogonal
direction to the radial direction. Thus, the Einstein field
equations finally take the form
\begin{eqnarray}
\rho(r)&=&-p_r(r)=-L-E^2L_F=\frac{1}{4\pi}
\;\frac{m'}{r^2}  \label{rho2}\,,\\
p_t(r)&=& L-\frac{1}{r^4\sin^2\theta}\,B^2\,L_{F}=-\frac{1}{8\pi}
\;\frac{m''}{r} \label{pr2}\,.
\end{eqnarray}
We will consider $|E^2\,L_{F}|<\infty$ and
$|B^2\,L_{F}/(r^4\sin^2\theta)|<\infty$, as $r\rightarrow 0$, to
ensure the regularity of the stress-energy tensor components.

One may now define the factor $\alpha'(r)$ as the {\it gravity
profile}, as it is related to the radial component of proper
acceleration that an observer must maintain in order to remain at
rest at constant $r$, $\theta$, $\phi$. Note that the radial
component of proper acceleration is given by
$a^r=(1-2m(r)/r)\,\alpha'(r)$ (see Ref. \cite{VDW} for details).
Thus, the convention used is that $\alpha'(r)$ is positive for an
inwardly gravitational attraction, and negative for an outward
gravitational repulsion \cite{VDW}. For gravastars an essential
condition is that they have a repulsive nature, i.e., $\alpha'<0$,
which for the present case amounts to imposing the following
condition
\begin{equation}
\alpha'(r)=\frac{m(r)-m'(r)r}{r[r-2m(r)]}<0  \,.
      \label{gravprofile}
\end{equation}
We emphasize that what is required from a spherically symmetric
solution of the Einstein equations to be a gravastar model, is the
presence of a repulsive nature of the interior solution, which is
characterized by the notion of the ``gravity profile''. The
interior solution is then matched to an exterior Schwarzschild
solution which is outlined further ahead.

We also explore the energy conditions, in particular, the weak
energy condition (WEC), which is defined as
$T_{\hat\mu\hat\nu}U^{\hat\mu}U^{\hat\nu}\geq0$ where
$U^{\hat\mu}$ is a timelike vector. The fact that the stress
energy tensor is diagonal will be helpful, so that the following
three conditions are imposed
\begin{equation}\label{WEC}
\rho \geq 0 \,,  \qquad  \rho+p_r \geq 0 \,, \qquad  \rho+p_t \geq
0  \,.
\end{equation}
From the first inequality, we verify that $m'\geq 0$ is imposed.
Note that $\rho+p_r=0$, so that the second inequality is readily
satisfied. Relatively to the third inequality, consider the factor
\begin{eqnarray}\label{WEC2}
\rho+p_t=-\left(E^2+\frac{1}{r^4\sin^2\theta}\,B^2\right)\;L_F
=\frac{N(r)}{8\pi r^2}  \,,
\end{eqnarray}
where for simplicity, $N(r)$ is defined as
\begin{eqnarray}
N(r)=2m'-rm'' \,.
 \label{def:N}
\end{eqnarray}
Thus, in summary, to ensure the WEC, we simply need to impose that
\begin{equation}\label{summaryWEC}
m'\geq 0   \qquad  {\rm and} \qquad  N(r) \geq 0
\end{equation}
throughout the spacetime.

In the present case, the term $\rho+p_t$ is equivalent to the
anisotropy factor which is defined as $\Delta=p_t-p_r$. The latter
is a measure of the pressure anisotropy of the fluid comprising
the gravastar. $\Delta=0$ corresponds to the particular case of an
isotropic pressure gravastar, which in our case reduces to the
trivial case of $E=B=0$, as may be verified from equation
(\ref{WEC2}). Note that $\Delta/r$ represents a force due to the
anisotropic nature of the stellar model, which is repulsive, i.e.,
being outward directed if $p_t>p_r$, and attractive if $p_t<p_r$.

We consider a cut-off of the stress-energy tensor at a junction
radius $a$, much in the spirit of the original Mazur-Mottola
gravastar model \cite{Mazur}. For instance, consider for
simplicity that the exterior solution is the Schwarzschild
spacetime, given by
\begin{eqnarray}
ds^2&=&-\left(1-\frac{2M}{r}\right)\,dt^2+
\left(1-\frac{2M}{r}\right)^{-1}dr^2
     +r^2(d\theta ^2+\sin ^2{\theta}\, d\phi ^2)
\label{metricvacuumlambda}  \,.
\end{eqnarray}
$M$ may be interpreted as the gravastar's total mass. In this case
the spacetimes given by the metrics (\ref{generalmetric}) and
(\ref{metricvacuumlambda}) are matched at $a$, and one has a thin
shell surrounding the gravastar. Note that to avoid the presence
of an event horizon, we need to impose $a>2M$. Using the
Darmois-Israel formalism, the surface stresses are given by
\begin{eqnarray}
\sigma&=&-\frac{1}{4\pi a} \left(\sqrt{1-\frac{2M}{a}+\dot{a}^2}-
\sqrt{1-\frac{2m}{a}+\dot{a}^2} \, \right)
    \label{surfenergy}   ,\\
{\cal P}&=&\frac{1}{8\pi a} \left(\frac{1-\frac{M}{a}
+\dot{a}^2+a\ddot{a}}{\sqrt{1-\frac{2M}{a}+\dot{a}^2}}
     -\frac{1-\frac{m}{a}-m'+\dot{a}^2
     +a\ddot{a}}{\sqrt{1-\frac{2m}{a}+\dot{a}^2}}
\, \right)         \,,
    \label{surfpressure}
\end{eqnarray}
where the overdot denotes a derivative with respect to the proper
time, $\tau$, and the prime represents, for the present case, a
derivative with respect to $a$. $\sigma$ is the surface energy
density and ${\cal P}$ the surface pressure (see Refs.
\cite{thinshell} for details). The static case is given by taking
into account $\dot{a}=\ddot{a}=0$. The total mass of the
gravastar, for the static case, with the junction interface $a_0$,
is given by
\begin{equation}\label{totalmass}
M=m(a_0)+m_s(a_0)\left[\sqrt{1-\frac{2m(a_0)}{a_0}}-\frac{m_s(a_0)}{2a_0}\right]
\,,
\end{equation}
where $m_s$ is the surface mass of the thin shell, and is defined
as $m_s=4\pi a^2 \sigma$.

It is also important to mention that for a self-gravitating object
(solution to Einstein's equations) to be considered as a possible
alternative to a black hole, which was one of the original ideas
behind the gravastar model, the surface redshift should be able to
reach values that are higher than that of ordinary objects. The
surface redshift is defined by $Z=\Delta
\lambda/\lambda_e=\lambda_0/\lambda_e-1$, where $\Delta\lambda$ is
the fractional change between the observed wavelength,
$\lambda_0$, and the emitted wavelength, $\lambda_e$. Thus,
according to our notation, the surface value takes the form
\begin{equation}
Z_{a_0}=e^{-\alpha(a_0)}-1 \,.
      \label{redshift}
\end{equation}
Now, for a static perfect fluid sphere the surface redshift is not
larger than $Z=2$ \cite{Buchdahl}. However, for anisotropic
spheres this value may be larger \cite{Ivanov}. For the present
case of nonlinear electrodynamic gravastars, the bounds on the
surface redshift, in principle, place restrictions on the
characteristic parameters of the nonlinear electrodynamic theory,
which can be analyzed on a case by case, but we will not pursue
this here.

\section{$F$ representation of nonlinear
electrodynamics}\label{sec2}

A number of nonexistence theorems prohibiting regular electrically
charged nonlinear electrodynamic structures have been proposed.
These basically state that any $L$ approaching the Maxwell weak
field limit, $F \ll 1$, such that $L \sim -F$, prohibit
electrically charged static spherically symmetric geometries with
a regular center \cite{Bronnikov1}. However, it has been argued
that due to the fact that the energy density attains a maximum at
the center, the validity of the Maxwell field limit at the center
cannot be expected \cite{Dymnikova2}. Another important feature
worth mentioning, is that imposing the Maxwell limit at the center
and at infinity, inevitably leads to a branching of $L$ as a
function of $F$. However, this issue should not bother us, as in
this paper we consider matchings of interior nonlinear
electrodynamic solutions to an exterior Schwarzschild spacetime,
and thus the range considered generally implies a monotonic
behavior of $L$ as a function of $F$.
As the Maxwellian limit is not imposed at the center, but rather
at infinity, the nonlinear electrodynamic constructions outlined
in this work do not possess the weak field limit in the specific
range of interest, i.e., $0 \leq r \leq a$, so that we are at a
certain freedom to consider rather general Lagrangians, in
particular, those that do not obey the Maxwellian limit for $F\ll
1$.

Now, taking into account the electromagnetic field equations,
equations (\ref{em-field}), from
$\left(^*F^{\mu\nu}\right)_{;\mu}=0$, we obtain $B'=0$ and
$E_{,\theta}=0$, so that we have $E=E(r)$ and $B=B(\theta)$.
From $\left(F^{\mu\nu}\,L_{F}\right)_{;\mu}=0$, we deduce
\begin{equation}\label{ELF}
EL_F=\frac{q_e}{r^2} \,,  \qquad  B=q_m\sin\theta  \,.
\end{equation}
The invariant $F=F^{\mu\nu}F_{\mu\nu}/4$, included for
self-completeness, takes the following form
\begin{equation}
F=-\frac{1}{2}\left(E^2-\frac{B^2}{r^4
\sin^2\theta}\right)=-\frac{1}{2}\left(E^2-\frac{q_m^2}{r^4}\right)
\,. \label{invF}
\end{equation}

Note that from the imposition of the WEC, equation (\ref{WEC2})
reduces to
\begin{eqnarray}\label{EMWEC2}
\rho+p_t=-\left(E^2+\frac{q_m^2}{r^4}\right)\;L_F
=\frac{N(r)}{8\pi r^2}  \,,
\end{eqnarray}
from which we verify $L_F\leq 0$. Now, from equations (\ref{ELF})
and (\ref{EMWEC2}), we determine the electric field, $E$, given by
\begin{equation}\label{E-field}
E(r)=\frac{1}{16\pi q_e}\left[-N(r) \pm
\sqrt{N^2(r)-\left(\frac{16\pi q_e q_m}{r^2}\right)^2}\right]
\,.
\end{equation}
One verifies that independently of $N(r)$, the electric field
diverges at the center. Thus, taking into account the {\it
principle of finiteness}, which states that a satisfactory theory
should avoid physical quantities becoming infinite, the message
that one may extract is that for the present geometry, both
electric and magnetic fields cannot coexist. One should either
consider a pure electric field or a pure magnetic field.
We emphasize that the principle of finiteness, which is a basic
requisite of nonlinear electrodynamic theory, is related to
physically measured entities or physical quantities, independently
of the particular basis, as introduced by Born and Infeld and
later retaken by Plabanski, such as the electric and magnetic
fields. The $F$ is a construct of the formalism, and therefore not
the physically measured entity.

\subsection{Pure electric field}

Considering a pure electric field, $B=0$, and the non-trivial case
of the negative sign in equation (\ref{E-field}), we have
\begin{equation}\label{pureE-field}
E(r)=-\frac{N(r)}{8\pi q_e} \,, \qquad F=-E^2/2=-\frac{N^2}{2(8\pi
q_e)^2}\,.
\end{equation}
From $EL_F=q_e/r^2$, the following relationship is deduced
\begin{equation}\label{LF-E}
L_F=-\frac{8\pi q_e^2}{Nr^2} \,,
\end{equation}
and the Lagrangian is provided by equation (\ref{pr2}), i.e.,
\begin{equation}\label{LagE}
L=-\frac{1}{8\pi} \frac{m''}{r}\,.
\end{equation}

Relatively to the WEC, one may consider a brief analysis by taking
into account the nonlinear electrodynamic quantities. From the
condition $\rho \geq 0$, we deduce $L \leq 2FL_F$, and as noted
above, $\rho+p_r=0$ is readily verified. From the condition
$\rho+p_t \geq 0$, we have $FL_F \geq 0$, and as $F<0$, then we
verify that $L_F\leq 0$.

\subsection{Pure magnetic field}

For a pure magnetic field, $E=0$, we have the following relevant
factors
\begin{equation}\label{pureB-field}
B=q_m\sin\theta\,, \qquad  F=\frac{q_m^2}{2r^4}     \,,
\end{equation}
and
\begin{equation}\label{LF-B}
L_F=-\frac{Nr^2}{8\pi q_m^2} \,.
\end{equation}
Using equation (\ref{rho2}), the Lagrangian is given by
\begin{equation}\label{LagB}
L=-\frac{1}{4\pi} \frac{m'}{r^2}\,.
\end{equation}
From the above equations, namely, from $F$ and $L$, one verifies
that specifying a convenient nonlinear electrodynamic Lagrangian,
equation (\ref{LagB}) may be integrated to provide the mass
function, and thus specifying the geometry of the solution. We
explore several cases in some detail in the next section.

Analogously to the previous case, one may analyze the WEC by
taking into account the nonlinear electrodynamic quantities. From
the inequality $\rho \geq 0$, we have $L \leq 0$. From the
condition $\rho+p_t \geq 0$, we deduce $-FL_F \geq 0$, and as
$F\geq 0$, then we verify that $L_F\leq 0$.

\subsubsection{Magnetic Dymnikova gravastar.}\label{sec:monopole}

In this section, we analyze the Lagrangian analogue of the
structural function, representing a regular electrically charged
structure, proposed by Dymnikova \cite{Dymnikova2}. Consider the
Lagrangian and its derivative given by
\begin{equation}\label{monopole}
L=-\frac{F}{4\pi\left(1+b\sqrt{2q_m^2F}\right)^2}\,, \qquad
L_F=-\frac{1}{4\pi\left(1+b\sqrt{2q_m^2F}\right)^3}\,,
\end{equation}
where $b$ is a characteristic parameter of the nonlinear
electrodynamic theory. Note that the stability of this solution
has also been analyzed in Ref. \cite{Breton}. In the weak field
limit, $F\ll 1$, the Lagrangian assumes the Einstein-Maxwell form,
$L\sim -F/(4\pi)$ and $L_F\sim -1/(4\pi)$.

Using $F=q_m^2/(2r^4)$, the above relationships take the following
form
\begin{equation}
L=-\frac{g^2}{8\pi(b g^2+r^2)^2}\,, \qquad L_F=-\frac{r^6}{4\pi (b
g^2+r^2)^3}\,,
\end{equation}
where we consider the definition $g=|q_m|$.

Now, using equation (\ref{LagB}), one may deduce the mass
function, which is given by
\begin{equation}\label{massmonopole}
m(r)=-\frac{g^2r}{4\left(b g^2+r^2\right)}
+\frac{g}{4\sqrt{b}}\,\arctan\left(\frac{r}{\sqrt{b}g}\right)\,.
\end{equation}
The stress-energy tensor components are given by
\begin{eqnarray}
\rho =-p_r=\frac{g^2}{8\pi(b g^2+r^2)^2} \qquad
p_t=\frac{g^2(r^2-b g^2)}{8\pi(b g^2+r^2)^3} \,,
\end{eqnarray}
which are regular throughout the spacetime. The WEC is satisfied
as can be verified from the following relationships
\begin{equation}
m'(r)=\frac{g^2r^2}{2(b g^2+r^2)^2}\,,  \qquad
N(r)=\frac{2g^2r^4}{(b g^2+r^2)^3}\,,
\end{equation}
which are manifestly positive.

\begin{figure}[h]
\centering
  \includegraphics[width=2.8in]{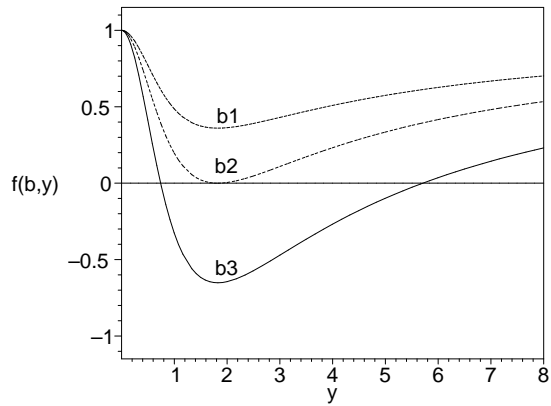}
  \caption{Roots of $f(b,y)$, with $b_1=0.2775$, $b_2=0.1775$ and
  $b_3=0.1075$, respectively. Note the presence of two positive
  roots, for $b<b_r\simeq 0.1775$, reflecting the existence of
  event horizons. For $b>b_r$, we have $f>0$.}
  \label{Fig:roots}
\end{figure}

The metric fields are given by
\begin{equation}
e^{2\alpha}=e^{-2\beta}=1-\left[-\frac{g^2}{2\left(b
g^2+r^2\right)}+\frac{g}{2\sqrt{b}\,r}\,\arctan\left(\frac{r}{\sqrt{b}g}\right)\right]\,.
\end{equation}
Defining $y=r/(\sqrt{b}g)$, the above equation provides
\begin{equation}
f(b,y)=1-\frac{1}{2b y }\left[\arctan(y)-\frac{y}{1+y^2}\right]\,.
\end{equation}
Using $\partial f/\partial y=0$, we verify that $f$ possesses a
single minimum, $y_m \approx 1.825$, independently of the value of
$b$. Now, $f(b,y_m)=0$ has a single positive root at $b \approx
0.1775$. For $b>b_r$, we have $f(b,y)>0$. If $b<b_r$, then
$f(b,y)$ possesses two roots, $y_1$ and $y_2$, which reflects the
existence of event horizons. Note that $f(b,y)>0$ for $y<y_1$ and
$y>y_2$, which are the cases we are interested in. This analysis
is represented in figure \ref{Fig:roots}.

\begin{figure}[h]
\centering
  \includegraphics[width=3.0in]{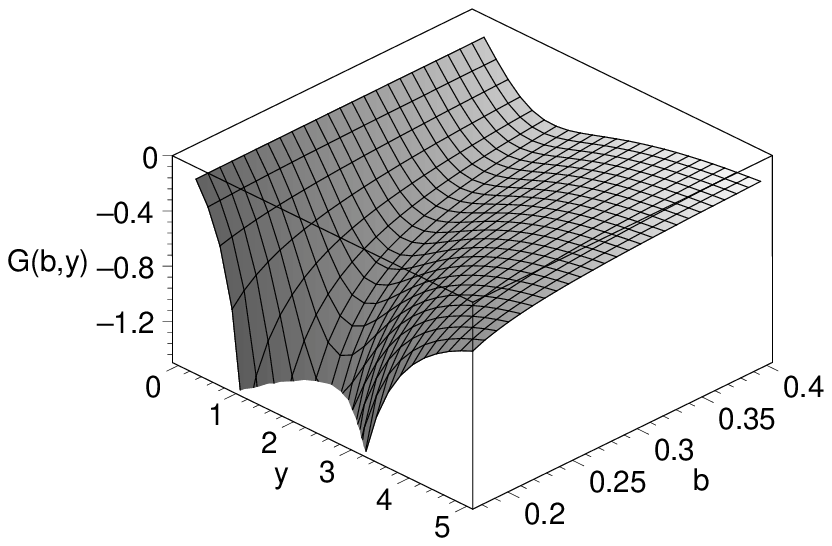}
  \hspace{0.1in}
  \includegraphics[width=2.4in]{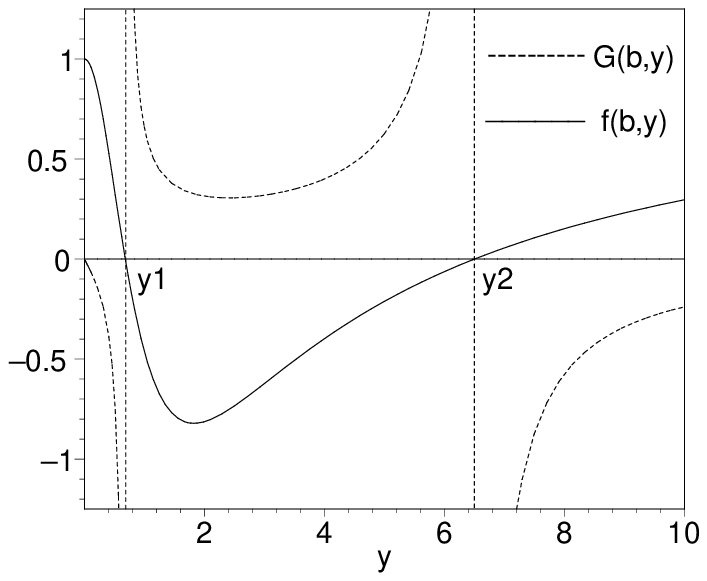}
  \caption{Plots of the gravity profile $G(b,y)=g\,
b^{3/2}\alpha'(b,y)$. In the left
  plot, we have considered $b>b_r$, and have shown that
  $\alpha'<0$ for all values of $y$. In the right plot,
  we have the case of $b<b_r$ and for $f>0$ the gravity
  profile is only negative in the range $y<y_1$ and $y>y_2$. We have
  considered the specific case of $b=0.0975$. The dashed curves
  represent the gravity profile and the solid line depicts
  $f(b,y)$.}
  \label{Fig:gravprofile-monople}
\end{figure}

Consider the gravity profile written in the following form
\begin{equation}
G(b,y)=g\,
b^{3/2}\alpha'(b,y)=-\frac{\frac{y(1+3y^2)}{1+y^2}-\arctan(y)}{4y^2\left\{
1-\frac{1}{2b y }\left[\arctan(y)-\frac{y}{1+y^2}\right]
\right\}}\,,
\end{equation}
which is depicted in figure \ref{Fig:gravprofile-monople}. For
$b>b_r$, we verify that $\alpha'<0$, i.e., the gravity profile is
negative, reflecting a repulsive character of the geometry, for
all values of $y$. For the case of $b<b_r$ and for $f>0$, the
gravity profile is only negative in the range $y<y_1$ and $y>y_2$.
In the right plot of figure \ref{Fig:gravprofile-monople}, we have
considered the specific case of $b=0.0975$.

Therefore, one may construct a nonlinear electrodynamic magnetic
gravastar, by matching the solution outlined above to an exterior
Schwarzschild solution. Note, however, that the magnetic
permeability, $\mu \sim 1/L_F$, diverges at the center, $r=0$, as
$L_F \rightarrow 0$, thus, exhibiting a magnetically
superconductive behavior. For $b>b_r$, we verify the absence of
event horizons and the gravity profile is negative for all values
of $y$, so that the matching occurs simply at $a>2M$. If $b<b_r$,
then more care needs to be taken due to the presence of two
horizons, at $r_1=\sqrt{b}\,gy_1$ and $r_2=\sqrt{b}\,gy_2$,
respectively. The gravity profile is negative for $r<r_1$, so that
the matching should be at $r_1>a_0>2M$, in order to avoid the
presence of event horizons. The surface stresses of the thin shell
are given by
\begin{eqnarray}
\fl \sigma&=&-\frac{1}{4\pi a_0} \left\{\sqrt{1-\frac{2M}{a_0}}-
\sqrt{1-\left[-\frac{g^2}{2\left(b
g^2+a_0^2\right)}+\frac{g}{2\sqrt{b}\,a_0}\,\arctan\left(\frac{a_0}
{\sqrt{b}g}\right)\right]} \, \right\}  \,,
    \label{magnetic-surfenergy}   \\
\fl {\cal P}&=&\frac{1}{8\pi a_0}
\left\{\frac{1-\frac{M}{a_0}}{\sqrt{1-\frac{2M}{a_0}}}
     -\frac{1-\frac{g^2(a_0^2-bg^2)}{4\left(b
g^2+a_0^2\right)}-\frac{g}{4\sqrt{b}\,a_0}\,\arctan\left(\frac{a_0}
{\sqrt{b}g}\right)}{\sqrt{1-\left[-\frac{g^2}{2\left(b
g^2+a_0^2\right)}+\frac{g}{2\sqrt{b}\,a_0}\,\arctan\left(\frac{a_0}
{\sqrt{b}g}\right)\right]} \,}\right\} \,.
    \label{magnetic-surfpressure}
\end{eqnarray}
The total mass of the magnetic monopole gravastar is provided by
the following relationship
\begin{eqnarray}
M&=&-\frac{g^2a_0}{4\left(b g^2+a_0^2\right)}
+\frac{g}{4\sqrt{b}}\,\arctan\left(\frac{a_0}{\sqrt{b}g}\right)
     \nonumber    \\
&&\hspace{-1.5cm}+m_s(a_0)\left[\sqrt{1+\frac{g^2}{2\left(b
g^2+a_0^2\right)}-\frac{g}{2\sqrt{b}\,a_0}\,
\arctan\left(\frac{a_0}{\sqrt{b}g}\right)}-\frac{m_s(a_0)}{2a_0}\right]
\,.
\end{eqnarray}

\subsubsection{Magnetic Bardeen gravastar.}

Bardeen seems to have been the first author to surprisingly
produce a regular black hole model \cite{Bardeen}. It is
interesting to note that the Bardeen model has been reinterpreted
as a magnetic solution to the Einstein field equation coupled to
nonlinear electrodynamics \cite{Garcia4}, i.e., it corresponds to
a self-gravitating magnetic monopole. (Gravitational magnetic
monopole stellar solutions have also been explored in
Majumdar-Papapetrou systems \cite{Zanchin}).

Consider the Lagrangian and its derivative, $L_F$, written in the
following form
\begin{equation}\label{Bardeen}
\fl L=-\frac{3}{4\pi
sq_m^2}\left(\frac{\sqrt{2q_m^2F}}{1+\sqrt{2q_m^2F}}\right)^{5/2}\,,
\qquad
L_F=-{\frac {15}{8\pi
s}}\,\frac{(2{q_m}^{2}F)^{1/4}}{(1+\sqrt{2{q_m}^2F}\,)^{7/2}}
 \,.
\end{equation}
Note, however, that this Lagrangian does not assume the Maxwell
form in the weak field limit, i.e., $L\sim -F F^{1/4}$ for $F\ll
1$.
It is perhaps important to emphasize here that to be a nonlinear
electrodynamic model, the Maxwellian limit has to be recovered in
the weak field limit. The nonexistence theorems, proposed in Ref.
\cite{Bronnikov1}, of electrically charged regular structures
imposed the weak field limit as $r \rightarrow 0$. However, it was
argued that as the energy density attains a maximum at the center,
the Maxwellian limit cannot be expected at $r=0$
\cite{Dymnikova2}. Note that it is only at $r\rightarrow \infty$
that the weak field limit is recovered $F\rightarrow 0$, for the
spacetimes considered in this work. Therefore, in the gravastar
models that we construct, it is not necessary to regain the weak
field limit in the specified range $0\leq r \leq a$, as the
interior nonlinear electrodynamic solution is matched at a
junction interface $a$, to an exterior vacuum geometry.

Using $F=q_m^2/(2r^4)$, then the above relationships take the form
\begin{equation}
L=-{\frac {3 \,g_B^{3}}{4\pi s\left ({r}^{2}+g_B^{2}\right )^{5/2}
}}\,, \qquad
L_F=-{\frac {15 \,g_B{r}^{6}}{8\pi s\left ({r}^{2}+g_B^{2}\right
)^{7/2 }}}
 \,,
\end{equation}
where we consider the definition $g_B=|q_m|$.

Using equation (\ref{LagB}), one may deduce the following mass
function, given by
\begin{equation}
m(r)={\frac {g_Br^{3}}{s\left ({r}^{2}+g_B^{2}\right )^{3/2}}}
 \,.
\end{equation}
The stress-energy tensor components are given by
\begin{equation}
\rho =-p_r= {\frac {3 \,g_B^{3}}{4\pi s\left
({r}^{2}+g_B^{2}\right )^{5/2} }}\,, \qquad
p_t={\frac {3 \,g_B^{3}\left (3\,{r}^{2}-2\,g_B^{2}\right )}{8\pi
s \left ({r}^{2}+g_B^{2}\right )^{7/2}}} \,.
\end{equation}
The WEC is satisfied, as may be verified from the following
relationships
\begin{equation}
m'(r)={\frac {3 \,g_B^{3}{r}^{2}}{s\left ({r}^{2}+g_B^{2}\right
)^{5/2} }} \,,   \qquad
N(r)={\frac {15 \,g_B^{3}{r}^{4}}{s\left ({r}^{2}+g_B^{2}\right
)^{7/2 }}}
 \,.
\end{equation}

The metric fields given by $f(r)=1-2m(r)/r$, may be rewritten as
\begin{equation}
f(s,y)=1-\frac{2y^2}{s(1+y^2)^{3/2}}
 \,,
\end{equation}
using the definition $y=r/g_B$. Taking into account $\partial
f/\partial y=0$, we verify that $f$ possesses a single minimum,
$y_m =\sqrt{2}$, independently of the value of $s$. Now,
$f(s,y_m)=0$ has a single positive root at $s_r \approx 0.7698$.
For $s>s_r$, we have $f(s,y)>0$. If $s<s_r$, then $f(s,y)$
possesses two positive roots, $y_1$ and $y_2$, corresponding to
two event horizons. We verify $f(s,y)>0$, for $y<y_1$ and $y>y_2$,
which are the cases that we are interested in.

\begin{figure}[h]
\centering
  \includegraphics[width=3.0in]{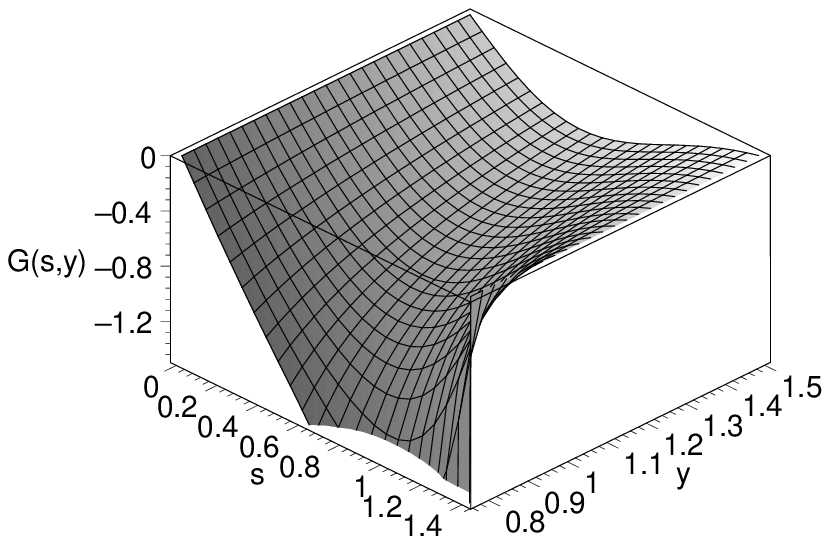}
  \hspace{0.1in}
  \includegraphics[width=2.5in]{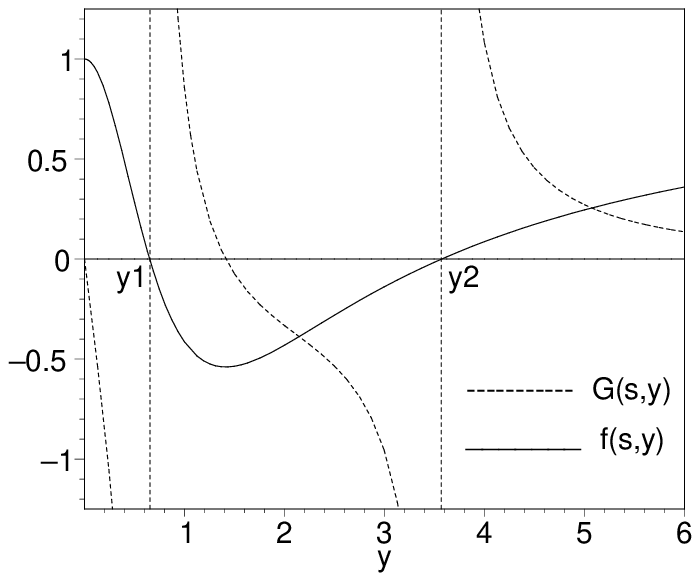}
  \caption{Plots of the gravity profile $G(s,y)=g_B\,\alpha'$,
  for the specific case of $s>s_r$ (left plot), and $s<s_r$ (right plot), respectively.
  For the first case,
  $s>s_r$, the gravity profile takes negative values,
  reflecting a repulsive character, in the range $0<y<y_m=
  \sqrt{2}$. For the second case, $s<s_r$ we have used $s=0.5$.
  The gravity profile is
  given by the dashed curves, and is negative in the range
  $y<y_1$, where $f>0$. The dashed curves represent the gravity
  profile, and the solid line depicts $f(s,y)$.}
  \label{Fig:Bardeengravprofile}
\end{figure}

The gravity profile given by
\begin{equation}
\alpha'=-\frac{g_Br(r^2-2g_B^2)}{(r^2+g_B^2)\left[2g_Br^2-s(r^2+g_B^2)^{3/2}\right]}
 \,,
\end{equation}
may be rewritten as
\begin{equation}
G=g_B\,\alpha'=-{\frac {y\left ({y}^{2}-2\right )}{\left
(1+{y}^{2}\right )\left (2\, {y}^{2}-s\left (1+{y}^{2}\right
)^{3/2}\right )}}  \,,
\end{equation}
using the definition $y=r/g_B$. The case of $s>s_r$, and for which
$\alpha'<0$, is plotted in figure \ref{Fig:Bardeengravprofile}.
Note that the gravity profile takes negative values, reflecting a
repulsive character, in the range $0<y<y_m$ for $s>s_r$. Thus, one
may match this interior solution at $r_m=g_B\,y_m>a_0>2M$. For the
case of $s<s_r$ and for $f>0$, the gravity profile is only
negative for $y<y_1$, being positive for $ y>y_2$. Note that the
gravity profile has asymptotes precisely at the roots of $f$.
Thus, one may match this solution at $r_1=g_B\,y_1>a_0>2M$, to an
exterior Schwarzschild spacetime.

The surface stresses of the thin shell are provided by
\begin{eqnarray}
\sigma&=&-\frac{1}{4\pi a_0} \left[\sqrt{1-\frac{2M}{a_0}}-
\sqrt{1-\frac{2g_Ba_0^2}{s(a_0^2+g_B^2)^{3/2}}} \, \right]
    \label{Bardeen-surfenergy}   ,\\
{\cal P}&=&\frac{1}{8\pi a_0}
\left\{\frac{1-\frac{M}{a_0}}{\sqrt{1-\frac{2M}{a_0}}}
     -\frac{1-\frac{g_Ba_0^2(a_0^2+4g_B^2)}{s(a_0^2+g_B^2)^{5/2}}}
     {\sqrt{1-\frac{2g_Ba_0^2}{s(a_0^2+g_B^2)^{3/2}}}
\,}\right\}
    \label{Bardeen-surfpressure}
\end{eqnarray}
The total mass of the Bardeen gravastar is given by
\begin{equation}
M={\frac {ga_0^{3}}{s\left ({a_0}^{2}+{g}^{2}\right
)^{3/2}}}+m_s(a_0)\left[\sqrt{1-{\frac {2ga_0^{2}}{s\left
({a_0}^{2}+{g}^{2}\right )^{3/2}}}}-\frac{m_s(a_0)}{2a_0}\right]
\,.
\end{equation}

\section{Dual $P$ formalism}\label{sec3}

Despite of the fact that nonlinear electrodynamics may be
represented in terms of a nonlinear electrodynamic field,
$F_{\mu\nu}$, and its invariants, one may introduce a dual
representation in terms of an auxiliary field $P_{\mu\nu}$. The
introduction of $P_{\mu\nu}$, proved to be extremely useful in the
derivation of exact solutions in general relativity, especially in
the electric regime. The dual representation of nonlinear
electrodynamics is obtained by a Legendre transformation given by
\begin{equation}\label{Hgen}
  H=2FL_F-L\,.
\end{equation}
The structural function $H$ is a function of a factor $P$, defined
by $P=P_{\mu\nu} P^{\mu\nu}/4$. In this representation, the theory
is reformulated in terms of the structural function $H$ by the
following relationships
\begin{eqnarray}
P_{\mu\nu}=L_F F_{\mu\nu}\,, \qquad F_{\mu\nu}=H_P P_{\mu\nu} \,,
\\
L=2PH_P-H\,, \qquad L_F H_P=1  \,,
\end{eqnarray}
where $H_P=dH/dP$. The invariant $P$ is given by
\begin{equation}\label{Pinv}
P=\frac{1}{4}\,P_{\mu\nu} P^{\mu\nu}=-\frac{1}{2}\left(P_{tr}^2
-\frac{1}{r^4\sin^2\theta}\,P_{\theta\phi}^2  \right)
\end{equation}

The stress-energy tensor in the dual $P$ formalism is written as
\begin{equation}
T_{\mu\nu}=g_{\mu\nu}\,(2PH_P-H)-P_{\mu\alpha}P_{\nu}{}^{\alpha}\,H_{P}\,,
\end{equation}
and in the orthonormal reference frame, provides the following
components
\begin{eqnarray}
T_{\hat{t}\hat{t}}&=&-T_{\hat{r}\hat{r}}=H-\frac{1}{r^4\sin^2\theta}\,P_{\theta\phi}^2\,H_P
\,,
     \label{4TttP}   \\
T_{\hat{\phi}\hat{\phi}}&=&T_{\hat{\theta}\hat{\theta}}
=-H-P_{tr}^2\,H_{P}  \,. \label{4TppP}
\end{eqnarray}

The electromagnetic field equations in the $P$ dual form are the
following
\begin{equation}
P^{\mu\nu}{}_{;\mu}=0  \;,  \qquad \left(H_P
\,^*P^{\mu\nu}\right)_{;\mu}=0 \,.
     \label{em-fieldP}
\end{equation}
We emphasize that the tensor $F_{\mu\nu}=H_P\,P_{\mu\nu}$ is the
physically relevant quantity. However, one may obtain electric
solutions easier by considering the $P$ dual formalism.

\subsection{Electric field}

Equation (\ref{Hgen}) implies that for the pure electric field,
$B=0$, together with equations (\ref{pureE-field})-(\ref{LagE})
provide the following extremely useful and simplified relationship
\begin{equation}\label{HpureE}
  H_E=\frac{1}{4\pi}\frac{m'}{r^2}\,,
\end{equation}
where $H_E$ is the $H$ structural function for this case and the
correspondent $P$ invariant
\begin{equation}\label{P-E}
  P_E=-\frac{q_e^2}{2r^4}\,.
\end{equation}
Thus, using the dual $P$ formalism, it is easier to find nonlinear
electrodynamic solutions than in the $F$ formalism, for the
specific case of pure electric fields. We consider several
solutions in this section.

\subsubsection{Ay\'{o}n-Beato$-$Garcia gravastar.}

In this section, we are interested in constructing a specific
gravastar geometry from a regular black hole solution coupled to
nonlinear electrodynamics, found by Ay\'{o}n-Beato$-$Garcia
\cite{Garcia2}. Although this solution is indeed regular, i.e.,
the metric, curvature invariants and the electric field are
regular everywhere, one still verifies the presence of event
horizons, and consequently the associated difficulties related to
these null hypersurfaces. Consider the specific
Ay\'{o}n-Beato$-$Garcia structural function \cite{Garcia2} given
in the following form
\begin{equation}\label{H-Bronnikov}
H=\frac{-P}{4\pi\cosh^2\left(s\sqrt[4]{-2q_e^2 P}\;\right)}\,,
\end{equation}
where $s$ is an adimensional constant. Note that $H$ assumes the
Einstein-Maxwell form in the weak field limit, i..e, $H\approx
-P/4\pi$ for $P\ll 1$. We also have
\begin{equation}\label{H_P-Bronnikov}
H_P=\frac{1}{8\pi\cosh^2\left(s\sqrt[4]{-2q_e^2
P}\;\right)}\;\left[s \,\tanh^2\left(s\sqrt[4]{-2q_e^2
P}\;\right)-2 \right]\,.
\end{equation}
Using equation (\ref{P-E}), $H$ and $H_P$ may be recast as
\begin{equation}
\fl H=\frac{q^2}{8\pi r^4\cosh^2\left(\frac{s q}{r}\right)}\,,
\qquad
H_P=\frac{1}{8\pi\cosh^2\left(\frac{s
q}{r}\right)}\;\left[\left(\frac{s q}{r}\right)
\,\tanh^2\left(\frac{s q}{r}\right)-2 \right]\,,
\end{equation}
where the definition $q=|q_e|$ is used.

The mass function is determined by using equation (\ref{HpureE}),
and is given by
\begin{equation}\label{M-Bronnikov}
m(r)=\frac{q}{2s}\left[1-\tanh\left(\frac{s q}{r}\right)\right]\,.
\end{equation}
With this solution at hand, the energy density and the principal
pressures take the following form
\begin{eqnarray}
\rho &=&-p_r=\frac{q^2}{8\pi r^4}\left[1-\tanh^2\left(\frac{s
q}{r}\right)\right]  \,,  \\
p_t&=&-\frac{q^2}{8\pi r^4}\left[1-\tanh^2\left(\frac{s
q}{r}\right)\right]\,\left[\left(\frac{s
q}{r}\right)\tanh\left(\frac{s q}{r}\right)-1\right] \,.
\end{eqnarray}

Analysing the geometry of the solution, consider the metric fields
\begin{equation}
e^{2\alpha(r)}=e^{-2\beta(r)}=1-\frac{q}{s \,r
}\left[1-\tanh\left(\frac{s q}{r}\right)\right] \,.
\end{equation}
Defining $y=r/(s q)$, the above equation takes the form
\begin{equation}
f(s,y)=1-\frac{\left[1-\tanh\left(1/y\right)\right]}{s^2y} \,.
\end{equation}
Using $\partial f/\partial y=0$, which may be re-expressed as
$[1-\tanh(1/y)][y-1-\tanh(1/y)]=0$, we verify that $f$ possesses a
single minimum, $y_m \approx 1.564$, independently of the value of
$s$. Now, $f(s,y_m)=0$ has a single positive root at $s \approx
0.5277$. However, for this case, we have $f(s,y)>0$ if $s>s_r$. If
$s<s_r$, then $f(s,y)$ possesses two roots, $y_1$ and $y_2$, which
reflects the existence of event horizons.

\begin{figure}[h]
\centering
  \includegraphics[width=2.8in]{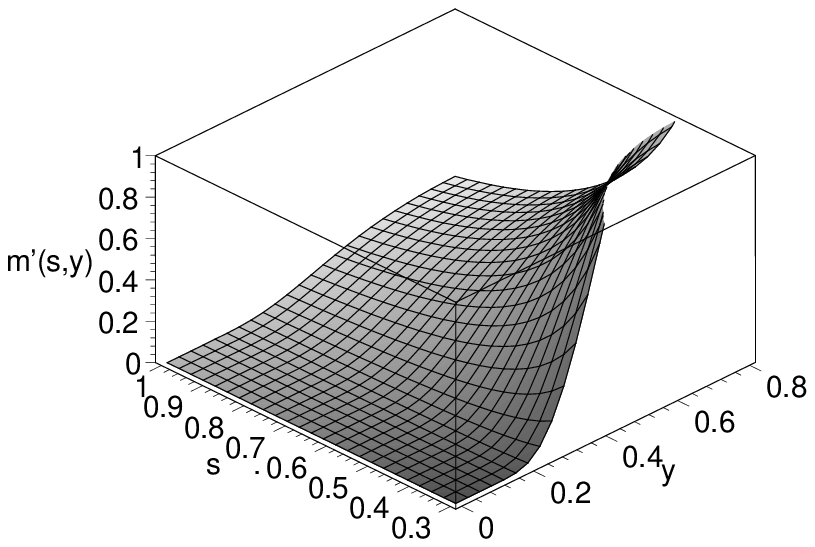}
  \includegraphics[width=2.8in]{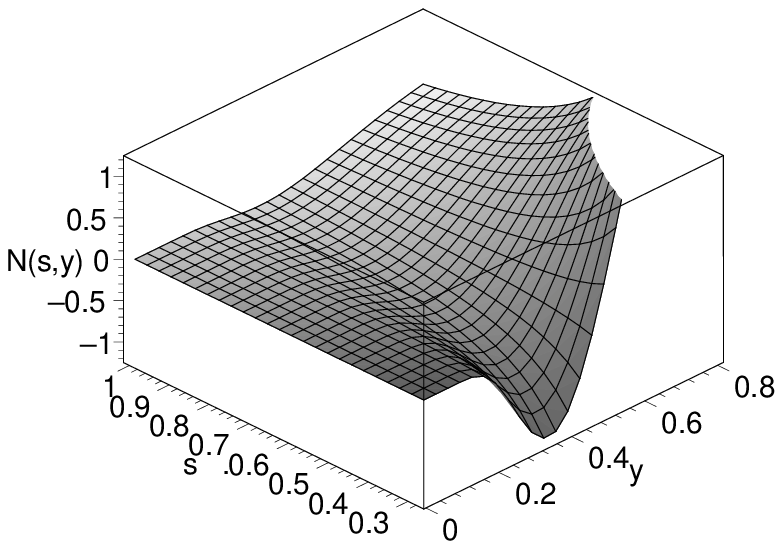}
  \caption{Plots of the WEC, $m'(r)$ and $N(r)$, respectively. We
  have only plotted the range $0.3<s<1$, for representational
  convenience. $m'\geq 0$ is verified for all values of $s$ and $y$.
  However, $N$ is negative in the range $0<y<y_N\simeq 0.4842$,
  for all values of $s$.}
  \label{Fig:WEC}
\end{figure}

We also need to verify the WEC, given by $m'\geq 0$ and $N\geq 0$,
which taking into account the definition $y=r/(s q)$, assume the
following form
\begin{eqnarray}
m'(s,y)&=&\frac{1}{2s^2y^2}\left[1-\tanh^2\left(\frac{1}{y}\right)\right] \,,   \\
N(s,y)&=&\frac{1}{s^2y^2}\left[1-\tanh^2\left(\frac{1}{y}\right)\right]
\,\left[2-\left(\frac{1}{y}\right)\tanh\left(\frac{1}{y}\right)\right]\,,
\end{eqnarray}
which are represented in figure \ref{Fig:WEC}. Note that at the
center $y=r=0$, the factor $N$ becomes zero, showing that
$p_r=p_t$, as was to be expected. We have only considered the
range $0.3<s<1$, for representational convenience. Note that
$m'\geq 0$ is verified for all values of $s$ and $y$. However, $N$
is negative in the range $0<y<y_N\simeq 0.4842$, for all values of
$s$. Thus the WEC is violated in the range $0<y<y_N\simeq 0.4842$.

\begin{figure}[h]
\centering
  \includegraphics[width=3.0in]{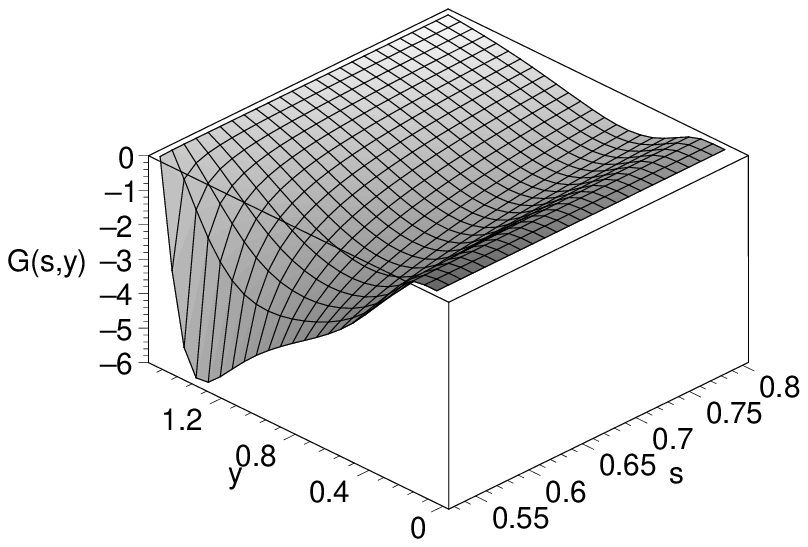}
  \hspace{0.1in}
  \includegraphics[width=2.5in]{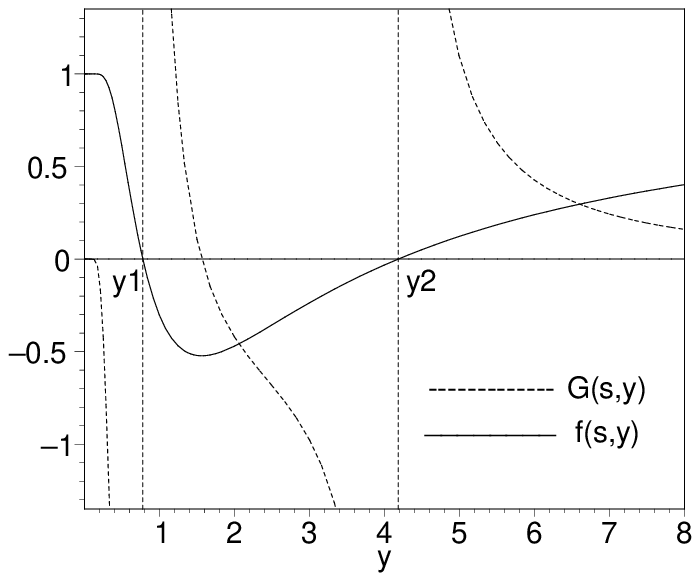}
  \caption{Plots of the gravity profile $G=(2s q)
  \alpha'$,
  for the specific case of $s>s_r$ (left plot), and $s<s_r$ (right plot), respectively.
  For the latter we have used $s=0.4277$. The gravity profile is
  given by the dashed curves. For the first case,
  $s>s_r$, the gravity profile takes negative values,
  reflecting a repulsive character, in the range $0<y<y_m\simeq 1.564$.
  For the case of $s<s_r$ and for $f>0$, the gravity
  profile is only negative for $y<y_1$.}
  \label{Fig:gravprofile}
\end{figure}

For the gravity profile we deduce the following relationship
\begin{equation}
\alpha'=\frac{\left\{1-\tanh\left(\frac{s
q}{r}\right)-\left(\frac{s
q}{r}\right)\left[1-\tanh^2\left(\frac{s
q}{r}\right)\right]\right\}}{2r\left[\frac{s
r}{q}-1+\tanh\left(\frac{s q}{r}\right)\right]}  \,,
\end{equation}
which using the definition $y=r/(s q)$, takes the form
\begin{equation}
G=(2s q)
\alpha'=\frac{\left\{1-\tanh\left(\frac{1}{y}\right)-\left(\frac{1}{y}\right)
\left[1-\tanh^2\left(\frac{1}{y}\right)\right]\right\}}
{y\left[s^2y-1+\tanh\left(\frac{1}{y}\right)\right]} \,.
\end{equation}
The case of $s>s_r$, and for which $\alpha'<0$, is plotted in
figure \ref{Fig:gravprofile}. Note that the gravity profile takes
negative values, reflecting a repulsive character, in the range
$0<y<y_m$ for $s>s_r$. Thus, one may match this interior solution
at $r_m=sqy_m>a_0>2M$. For the case of $s<s_r$ and for $f>0$, the
gravity profile is only negative for $y<y_1$, being positive for $
y>y_2$. Note that the gravity profile has asymptotes precisely at
the roots of $f$. Thus, one may match this solution at
$r_1=sqy_1>a_0>2M$, to an exterior Schwarzschild spacetime.

Then, with our solution for the mass function, alongside $H$ and
$H_P$, we can cast the following relevant functions for the system
\begin{equation}\label{KpureE-field}
E(r)=-\frac{q}{8\pi r^2} \left[1-\tanh^2\left(\frac{s
q}{r}\right)\right] \left[2-\left(\frac{s
q}{r}\right)\tanh\left(\frac{s q}{r}\right)\right] \,,
\end{equation}
\begin{equation}\label{K-F-E}
F=-\frac{1}{2}\left\{\frac{q}{8\pi r^2}
\left[1-\tanh^2\left(\frac{s q}{r}\right)\right]
\left[2-\left(\frac{s q}{r}\right)\tanh\left(\frac{s
q}{r}\right)\right]\right\}^2 \,,
\end{equation}
\begin{equation}\label{K-LF-E}
L_F=-8\pi\left\{ \left[1-\tanh^2\left(\frac{s q}{r}\right)\right]
\left[2-\left(\frac{s q}{r}\right)\tanh\left(\frac{s
q}{r}\right)\right] \right\}^{-1}
\end{equation}
\begin{equation}\label{KLagE}
L=-\frac{q^2}{8\pi r^4} \left[1-\tanh^2\left(\frac{s
q}{r}\right)\right] \left[\left(\frac{s
q}{r}\right)\tanh\left(\frac{s q}{r}\right)-1\right]\,,
\end{equation}

\begin{equation}\label{K-2FLF}
2FL_F=\frac{q^2}{8\pi r^4} \left[1-\tanh^2\left(\frac{s
q^2}{r^2}\right)\right] \left[2-\left(\frac{s
q}{r}\right)\tanh\left(\frac{s q}{r}\right)\right] \,.
\end{equation}

Given this, the above interior Ay\'{o}n-Beato$-$Garcia spacetime
may be matched to an exterior Schwarzschild solution, at the
junction interface, $a_0$, with the surface stresses given by
\begin{eqnarray}
\fl \sigma&=&-\frac{1}{4\pi a_0} \left(\sqrt{1-\frac{2M}{a_0}}-
\sqrt{1-\frac{q}{s a_0}\left[1-\tanh\left(\frac{s
q}{a_0}\right)\right]} \, \right)
    \label{ABGsurfenergy}   ,\\
\fl {\cal P}&=&\frac{1}{8\pi a_0}
\left[\frac{1-\frac{M}{a_0}}{\sqrt{1-\frac{2M}{a_0}}}
     -\frac{1-\left(\frac{q}{2s a_0}\right)\left\{1-\tanh\left(\frac{s
q}{a_0}\right)+\left(\frac{s
q}{a_0}\right)\left[1-\tanh^2\left(\frac{s
q}{a_0}\right)\right]\right\}}{\sqrt{1-\frac{q}{s
a_0}\left[1-\tanh\left(\frac{s q}{a_0}\right)\right]}} \, \right]
\,.
    \label{ABGsurfpressure}
\end{eqnarray}
The total mass of the gravastar is given by
\begin{equation}
\fl M=\left(\frac{q}{2s}\right)\left[1-\tanh\left(\frac{s
q}{a_0}\right)\right]+m_s(a_0)\left\{\sqrt{1-\frac{q}{s
a_0}\left[1-\tanh\left(\frac{s q}{a_0}\right)\right]}
-\frac{m_s(a_0)}{2a_0}\right\}\,,
\end{equation}
where $m_s(a_0)$ is the surface mass of the thin shell, given by
$m_s(a_0)=4\pi a_0^2 \sigma$.

Summarizing the above results, we have the following: $(i)$ For
$s>s_r$, the factor $f$ is positive for all values of $y$, proving
the absence of event horizons. However, the gravity profile,
$\alpha'$, is only negative in the range $0<y<y_m$, reflecting a
repulsive character of the geometry, which is essential for
gravastar solutions. Finally, the WEC is satisfied only in the
range $y\geq y_N$, for all values of $s$. Thus, in conclusion, one
may match this interior Ay\'{o}n-Beato$-$Garcia geometry to an
exterior Schwarzschild solution, at $r_m=sqy_m>a_0
>2M$, whilst the WEC is violated in the range $0<y<y_N<y_m$.
$(ii)$ For $s<s_r$, the factor $f$ has two roots, $y_1$ and $y_2$.
$f(s,y)$ is positive for $y<y_1$ and $y>y_2$, while the gravity
profile is only negative in the range $y<y_1$. As for the previous
case, the WEC is violated in the range $y<y_N<y_1$, as $N<0$.
Therefore, to construct a gravastar geometry, one needs to match
the Ay\'{o}n-Beato$-$Garcia solution to an exterior Schwarzschild
spacetime at $r_1=sqy_1>a_0>2M$, with the consequent violation of
the WEC in the range $0<y<y_N<y_1$. Another important
characteristic that both cases exhibit is an electrically
superconductive behavior, at the center, as $L_F\rightarrow\infty$
as $r\rightarrow 0$.

One may also consider other regular black hole solutions obtained
by Ay\'{o}n-Beato$-$Garcia, in particular the structural functions
given in Refs. \cite{Garcia,Garcia3}, however, we shall not
endeavor in this analysis. The message that one may extract, is
that one may, in principle, construct a wide variety of gravastar
models in the context of nonlinear electrodynamics, by using the
regular black hole solutions found by Ay\'{o}n-Beato$-$Garcia
\cite{Garcia,Garcia2,Garcia3}.

\subsubsection{Nonlinear modified Tolman-Matese-Whitman mass function.}

Consider the following structural function and its derivative
given by
\begin{equation}\label{1st}
H=\frac{\sqrt{-2P}(1+3\alpha\,\sqrt{-2P}\,)}{\alpha(1+\alpha\,\sqrt{-2P})^2}\,,
   \qquad
H_P=-\frac{(1+5\alpha\,\sqrt{-2P}\,)}{\alpha\sqrt{-2P}(1+\alpha\,\sqrt{-2P})^3}\,.
\end{equation}
Note, however, that this structural function does not assume the
Einstein-Maxwell form in the weak field limit, i..e, $H\approx
(-P)^{1/2}$ for $P\ll 1$. However, as mentioned above, we are not
preoccupied with regaining the weak field limit in the specified
range $0\leq r \leq a$, as the interior nonlinear electrodynamic
solution is matched at a junction interface $a$, to an exterior
vacuum geometry.

Taking into account $P=-q_e^2/(2r^4)$, $H$ and $H_P$ take the form
\begin{equation}
H=\frac{3+\frac{r^2}{\alpha q}}{\alpha^2\left(1+\frac{r^2}{\alpha
q}\right)^2}\,, \qquad
H_P= -\frac{\left(\frac{r^2}{\alpha
q}\right)^3\left(5+\frac{r^2}{\alpha
q}\right)}{\left(1+\frac{r^2}{\alpha q}\right)^3} \,,
\end{equation}
where the definition $q=|q_e|$ is used. The mass function may be
integrated to yield
\begin{equation}\label{m1st}
m(r)=\frac{4\pi r^3}{\alpha^2\left(1+\frac{r^2}{\alpha
q}\right)}\,.
\end{equation}
Note that this mass function is similar in form to the
Tolman-Matese-Whitman function analyzed in detail in Ref.
\cite{darkstar}, which represents a monotonic decreasing energy
density in the gravastar interior, and was used previously in the
analysis of isotropic fluid spheres by Matese and Whitman
\cite{MatWhit}. Thus, we denote the mass function given by
equation (\ref{m1st}) the nonlinear modified Tolman-Matese-Whitman
mass function.

The stress-energy tensor components take the following form
\begin{equation}
\rho=-p_r={\frac {q\left (3\alpha q+{r}^{2}\right )}{\alpha \left
(\alpha q +{r}^{2}\right )^{2}}}
 \;,
\qquad
p_t=-{\frac {{q}^{2}\left (3\alpha q-{r}^{2}\right )}{\left
(\alpha q+ {r}^{2}\right )^{3}}}
 \,.
\end{equation}
The WEC is also satisfied as the following factors
\begin{equation}
m'(r)={\frac {4\pi {r}^{2}q\left (3\alpha q+{r}^{2}\right
)}{\alpha \left (\alpha q+{r}^{2}\right )^{2}}} \;, \qquad
N(r)={\frac {8\pi {r}^{4}q\left (5\alpha q+{r}^{2}\right )}{\alpha
\left (\alpha q+{r}^{2}\right )^{3}}} \;,
\end{equation}
are manifestly positive. Note that at the center, $r=0$, we have
$N=0$, so that $p_r=p_t$, as was to be expected.

Analysing the geometry of the solution, consider
\begin{equation}
e^{2\alpha(r)}=e^{-2\beta(r)}=1-\frac{8\pi r^2}{\alpha^2
\left(1+\frac{r^2}{\alpha q} \right)} \,,
\end{equation}
Defining $s^2=8\pi q/\alpha$ and $y^2=r^2/(\alpha q)$, so that the
above equation takes the form
\begin{equation}
f(s,y)=1-\frac{s^2y^2}{1+y^2} \,.
\end{equation}
Note the absence of event horizons for $s<1$. If $s>1$, then $f$
possesses a positive root situated at $y_r=1/\sqrt{s^2-1}$.

\begin{figure}[h]
\centering
  \includegraphics[width=2.8in]{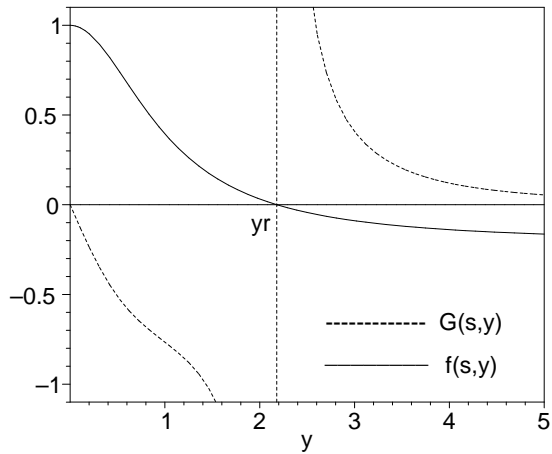}
  \caption{Plot of the gravity profile $G=\sqrt{\alpha q}\;\alpha'$,
  for the specific case of $s>s_r$.
  We have used $s=1.1$. The gravity profile takes negative values,
  reflecting a repulsive character, in the range $0<y<y_r$.}
  \label{Fig:gravprofileTMW}
\end{figure}

The gravity profile is given by
\begin{equation}
\alpha'=-{\frac {8\pi\alpha {q}^{2}r }{\left ({\alpha}^{2}q+\alpha
{r}^{2} -8\pi {r}^{2}q\right )\left (\alpha q+{r}^{2}\right )}}
\;.
\end{equation}
which using the definitions $s^2=8\pi q/\alpha$ and
$y^2=r^2/(\alpha q)$, takes the following form
\begin{equation}
G=\sqrt{\alpha q}\;\alpha'=-\frac{s^2
y}{\left[1+(1-s^2)y^2\right](1+y^2)}\;.
\end{equation}
One readily verifies that $\alpha'<$ for $s\leq 1$. For $s\geq 1$,
we also verify that $\alpha'<0$ for $y<y_r$. See figure
\ref{Fig:gravprofileTMW} for the latter case, with the specific
choice of $s=1.1$. An asymptote of $\alpha'$ exists precisely at
the root.

The electric field and the $F$ invariant are given by
\begin{equation}
E(r)=-{\frac {{r}^{4}\left (5\alpha q+{r}^{2}\right
)}{\alpha\,\left ( \alpha q+{r}^{2}\right )^{3}}} \,, \qquad
F=-\frac{1}{2}\left[{\frac {{r}^{4}\left (5\alpha q+{r}^{2}\right
)}{\alpha\left ( \alpha q+{r}^{2}\right )^{3}}} \right]^2 \,,
\end{equation}
respectively. We also have the following relationships
\begin{equation}
L=-{\frac {{q}^{2}\left (3\alpha q-{r}^{2}\right )}{\left (\alpha
q+ {r}^{2}\right )^{3}}}\;, \qquad
L_F=-{\frac {q\alpha \left (\alpha q+{r}^{2}\right
)^{3}}{{r}^{6}\left ( 5\alpha q+{r}^{2}\right )}} \,,
\end{equation}

In summary, one may match the interior solution to an exterior
Schwarzschild spacetime, at a junction interface, $a_0>2M$. If
$s>1$, then the matching needs to be done at $r_r=\sqrt{\alpha
q}\, y_r>a_0>2M$. The surface stresses are given by
\begin{eqnarray}
\fl \sigma&=&-\frac{1}{4\pi a_0} \left\{\sqrt{1-\frac{2M}{a_0}}-
\sqrt{1-\frac{8\pi a_0^2}{\alpha^2\left(1+ \frac{a_0^2}{\alpha
q}\right)}} \, \right\}
    \label{TMWsurfenergy}   ,\\
\fl {\cal P}&=&\frac{1}{8\pi a_0}
\left\{\frac{1-\frac{M}{a_0}}{\sqrt{1-\frac{2M}{a_0}}}
     -\left[1-\frac{8\pi a_0^2\left(2+ \frac{a_0^2}{\alpha
q}\right)}{\alpha^2\left(1+ \frac{a_0^2}{\alpha
q}\right)^2}\right]\Bigg/\sqrt{1-\frac{8\pi
a_0^2}{\alpha^2\left(1+ \frac{a_0^2}{\alpha q}\right)}} \,
\right\} \,.
    \label{TMWsurfpressure}
\end{eqnarray}
The total mass of the gravastar is given by
\begin{equation}
M=\frac{4\pi a_0^3}{\alpha^2\left(1+\frac{a_0^2}{\alpha
q}\right)}+m_s(a_0)\left\{\sqrt{1-\frac{8\pi
a_0^2}{\alpha^2\left(1+\frac{a_0^2}{\alpha q}\right)}}
-\frac{m_s(a_0)}{2a_0}\right\}\,,
\end{equation}
where $m_s(a_0)$ is the surface mass of the thin shell, given by
$m_s(a_0)=4\pi a_0^2 \sigma$.

\subsection{Magnetic field}

Now for the $E=0$ case, equations (\ref{pureB-field})-(\ref{LagB})
together with (\ref{Hgen}) provide the following relationships
\begin{equation}\label{HpureB}
  H_B=\frac{1}{8\pi}\frac{m''}{r}\,,
\end{equation}
where $H_B$ is the $H$ structural function for this case and the
correspondent $P$ invariant is given by
\begin{equation}\label{P-B}
  P_B=\frac{N^2}{2(8\pi q_m)^2}\,.
\end{equation}
We can, in principle, find new solutions for the mass function
$m(r)$ by choosing a suitable form of the $H=H_B$ structural
function. Although this treatment is far from being trivial, we
have considered the $F$ formalism to find purely magnetic
solutions.

\section{Summary and duscussion}\label{sec:conclusion}

Gravastar models have recently been proposed as an alternative to
black holes, mainly to avoid the associated difficulties with
event horizons and singularities. In this work, we have been
interested in the construction of gravastar models coupled to
nonlinear electrodynamics. Using the $F$ representation, we have
considered specific forms of Lagrangians describing magnetic
gravastars, which may be interpreted as self-gravitating magnetic
monopoles with a charge $g$. Using the $P$ formulation of
nonlinear electrodynamics, it is easier to find electric
solutions, and considering specific structural functions we
further explored the characteristics and physical properties of
these solutions. We have matched these interior nonlinear
electrodynamic geometries to an exterior Schwarzschild spacetime
at a junction interface $a$, thus avoiding the problematic issues
related to the singularities and event horizons.
We emphasize that what is required from a spherically symmetric
solution of the Einstein equations to be a gravastar model, is the
presence of a repulsive nature of the interior solution, which is
characterized by the notion of the ``gravity profile''.
It is important to point out that to be a nonlinear electrodynamic
model, the Maxwellian limit, $L\sim -F$ and $L_F\sim -1$, has to
be recovered in the weak field limit, $F\ll 1$. For the spacetimes
considered in this work, it is only at $r\rightarrow \infty$ that
the weak field limit, $F\rightarrow 0$, is recovered. Therefore,
in the gravastar models that we construct, it is not necessary to
regain the weak field limit in the specified range $0\leq r \leq
a$ of the interior nonlinear electrodynamic solution. Thus, we
have used general Lagrangians and structural functions that are
strongly non-Maxwellian in the weak field limit, namely the
Bardeen and the Tolman-Matese-Whitman solutions.
In conclusion, a rather wide variety of gravastar solutions may be
constructed within the context of nonlinear electrodynamics.



\end{document}